\documentclass{PoS}
\usepackage{graphicx,amsmath,amssymb,wrapfig}
\makeatletter
\newcommand*{\rom}[1]{\expandafter\@slowromancap\romannumeral #1@}
\makeatother

\usepackage{color}

\title{Theoretical perspective for the future experiments on parton densities}
\ShortTitle{Perspective for future experiments on PDFs}

\author{S. Kumano$^{\,\, a,b}$ \\
$^a$ KEK Theory Center, Institute of Particle and Nuclear Studies, KEK,\\
\ \ \            and Department of Particle and Nuclear Physics,
Graduate University for Advanced Studies \\
\ \ \  (SOKENDAI),       Ooho 1-1, Tsukuba, Ibaraki, 305-0801, Japan\\ 
$^b$ J-PARC Branch, KEK Theory Center,
     Institute of Particle and Nuclear Studies, KEK,\\
\ \ \ and Theory Group, Particle and Nuclear Physics Division, J-PARC Center,\\
\ \ \ 203-1, Shirakata, Tokai, Ibaraki, 319-1106, Japan\\
        E-mail: \email{shunzo.kumano@kek.jp}}

\abstract{I explain the current status of parton-distribution-function (PDF) 
studies and future experimental prospects on their determinations. 
First, unpolarized PDFs of the nucleon are introduced as a field of precision
QCD physics including higher-order $\alpha_s$ corrections. Second, nuclear PDFs and 
polarized nucleonic PDFs are discussed. Third, the determination of 
fragmentation functions is explained. Forth, the three-dimensional (3D)
structure functions are discussed in connection with the origin of
the nucleon spin and gravitational form factors of hadrons.
By the 3D structure functions, gravitational sources and the origin
of nucleon mass could be clarified in the microscopic level of quarks and gluons.
Furthermore, 3D structure studies of hadrons could be used 
for clarifying internal structure of exotic hadron candidates,
nuclear composition of ultra-high-energy cosmic rays,
and color-entanglement phenomena. The PDF field will be developed further 
along with progress in other fields of science.}

\FullConference{XXVI International Workshop on Deep-Inelastic Scattering 
        and Related Subjects (DIS2018)\\ 16-20 April 2018\\ Kobe, Japan}

\begin{document}

\section{Introduction}

Hadron physics is the field of science to study 
material creation in the universe and 
properties of quark-hadron many-body systems 
with ultimate densities in nature.
We know that the basic theory for strong interactions is 
Quantum Chromodynamics (QCD), which describes
quark and gluon interactions. 
Studies of the parton distribution functions 
(PDFs) of hadrons are a part of our efforts for universal 
understanding of the quark-hadron many-body systems 
from low to high densities, from low to high temperatures, 
and from low to high energies.

\begin{figure}[b!]
\vspace{0.10cm}
\begin{minipage}{\textwidth}
\begin{tabular}{lc}
\hspace{-0.40cm}
\begin{minipage}[c]{0.45\textwidth}
   \vspace{-0.0cm}
   \begin{center}
     \includegraphics[width=7.0cm]{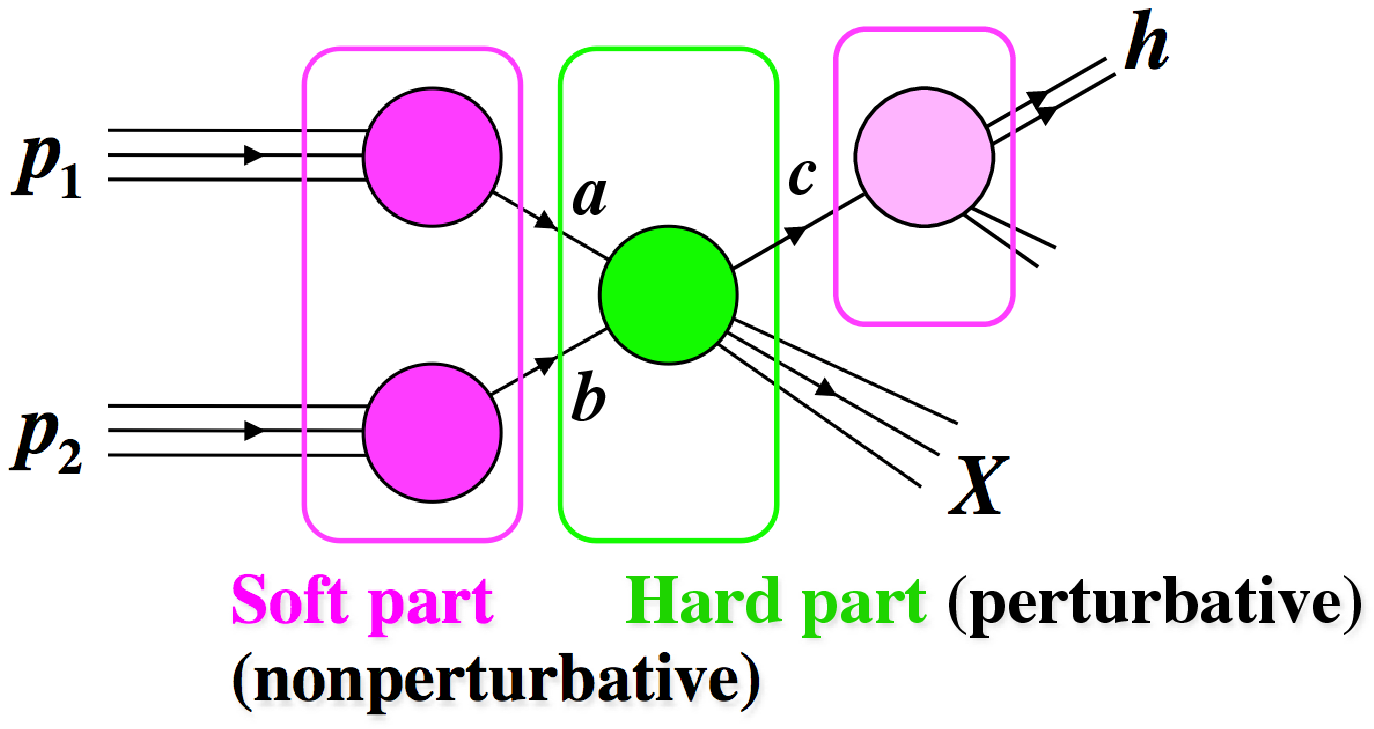}
   \end{center}
\vspace{-0.35cm}
\caption{High-energy hadron reactions.}
\label{fig:hard-cross}
\vspace{-0.4cm}
\end{minipage} 
\hspace{-0.0cm}
\begin{minipage}[c]{0.54\textwidth}
    \vspace{-0.5cm}
   \begin{center}
    \includegraphics[width=7.5cm]{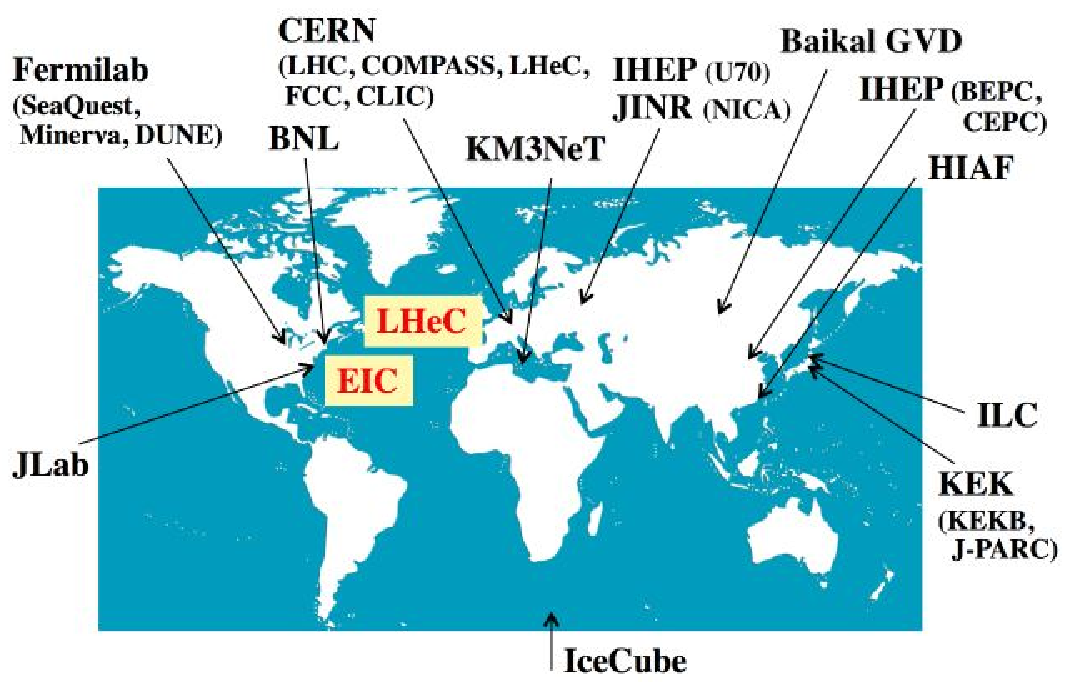}
   \end{center}
\vspace{-0.80cm}
\caption{Current and future facilities.}
\label{fig:facilities}
\vspace{-0.4cm}   
\end{minipage}
\end{tabular}
\vspace{0.20cm}
\end{minipage}
\end{figure}

The PDFs indicate parton momentum distributions 
inside a hadron, and they play an important role in calculating
high-energy hadron cross sections precisely as shown typically
in Fig.\,\ref{fig:hard-cross}.
Without their precise determination, it is impossible to find
any new physics beyond the current standard model and any new
phenomenon in high-energy lepton-hadron and hadron-hadron
reactions, for example, at LHC (Large Hadron Collider).
Thank to significant theoretical and experimental efforts 
on high-energy hadron reactions including higher-order 
$\alpha_s$ corrections, QCD became a field of precision physics, 
roughly speaking, within a several percent level, 
although the precision depends on kinematical regions 
as one will find later as uncertainty bands of the PDFs.

Recent PDF studies are not only on the traditional 
unpolarized nucleonic PDFs but also on nuclear PDFs, polarized PDFs,
fragmentation functions, and 3-dimensional structure functions.
As the nucleonic PDFs become more and more accurate, 
nuclear corrections cannot be ignored anymore if nuclear data
are partially included in the ``nucleonic'' PDF analysis.
The corrections are of the order of 10-30\% for medium and large nuclei. 
The nuclear PDFs are valuable in finding new phenomena 
in high-energy heavy-ion collisions and also
in high-precision neutrino oscillation experiments.
The polarized PDFs are essential for understanding the origin of nucleon
spin, which is one of fundamental physics quantities. 
We found that orbital-angular-momentum contributions of partons 
could become a significant part of the nucleon spin, and they should be
investigated by the three-dimensional structure functions such as
the generalized parton distributions (GPDs).
In addition to inclusive lepton and hadron reactions, semi-inclusive
processes, $e.g.$ pion production, become usual tools in 
investigating the internal structure of hadrons, such as 
flavor dependence of the PDFs and transverse-momentum-dependent 
parton distribution (TMDs), and quark-gluon plasma properties.
For descriptions of high-energy hadron productions, accurate
fragmentation functions are essential.

In this article, the current status and prospects are
discussed on these topics by considering future experimental projects.
For the DIS (deep inelastic scattering) community of this conference, 
the EIC (Electron-Ion Collider) and LHeC (Large Hadron Electron Collider)
are two major important projects in the middle of 2020's.
Here in Japan, there are projects of KEKB, 
J-PARC (Japan Proton Accelerator Research Complex), and
ILC (International Linear Collider) for structure-function 
and fragmentation-function studies as explained in this article.
In addition, there are ongoing and future world-wide projects at
Jefferson Lab 12 GeV, CERN (COMPASS, LHC, FCC, CLIC),
Fermilab (Sea-Quest, Miner$\nu$a, DUNE), BNL (RHIC), Chinese IHEP (BEPC, CEPC),
HIAF, Russian IHEP (U70), JINR (NICA), KM3NeT, and Baikal GVD
as shown in Fig.\,\ref{fig:facilities}.
In Sec.\,\ref{unpol-pdfs}, the unpolarized PDFs of the nucleon are
discussed, and their nuclear corrections are explained 
in Sec.\,\ref{nuclear-pdfs}. The polarized PDFs,
fragmentation functions, and 3D structure functions 
are discussed in Secs.\,\ref{polarized-pdfs}, \ref{ffs}, and 
\ref{3d-sfs}, respectively. These discussions are summarized 
in Sec.\ref{summary}.

\section{Unpolarized PDFs of the nucleon}
\label{unpol-pdfs}

\begin{wrapfigure}[13]{r}{0.30\textwidth}
   \vspace{-0.8cm} 
   \begin{center}
     \includegraphics[width=4.5cm]{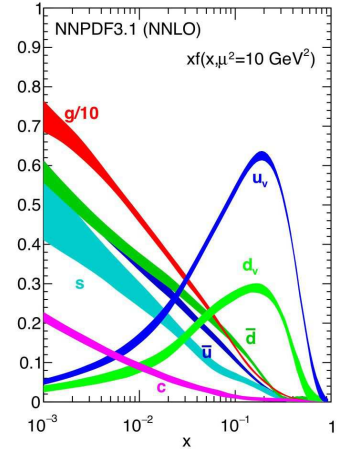}
   \end{center}
\vspace{-0.85cm}
\caption{Recent PDFs \cite{unpol-pdfs}.}
\label{fig:nnpdfs}
\vspace{-0.5cm}
\end{wrapfigure}

The unpolarized PDFs of the nucleon have been investigated 
for a long time, and this topic is considered as 
a precision QCD field. 
As the higher-order $\alpha_s$ corrections are now included
in the NNLO (next-to-next to leading order) level as a standard
\cite{unpol-pdfs}, and the N$^3$LO evolution and coefficient functions 
are within reach of theoretical studies \cite{n3lo}.
The PDFs are determined by a global $\chi^2$ analysis of high-energy 
hadron reaction data. They are expressed by a number of parameters
at a fixed $Q^2$, and they are evolved to experimental $Q^2$ points
to calculate $\chi^2$ by the DGLAP 
(Dokshitzer-Gribov-Lipatov-Altarelli-Parisi) evolution equations
including higher-order corrections. Recently, another analysis
of using a neutral-network method has been in progress, and 
it could be a reliable error estimate of the PDFs because 
it does not rely on a specific functional form. 

Recent typical PDFs are shown in Fig.\,\ref{fig:nnpdfs} with error bands.
Now, the distributions are well determined from small Bjorken-$x$ 
($\sim 10^{-3}$) to large $x$ in the NNLO. The strange-quark distribution
has been determined by the neutrino-induced opposite-sign dimuon 
measurements. Roughly, it is about 40\% of the light antiquark
distributions, $s \sim 0.4 (\bar u + \bar d)/2$ at small $Q^2$.
The CMS measurements on $W$ and charm production agree 
with this strange-quark distribution.
However, the ATLAS collaboration suggested that $s$ 
is similar to the light antiquark
distributions. There were discussions
in the parallel sessions of this conference including analysis method.
It may take a few yeas to resolve this issue.

As the PDFs become more and more accurate, we hope that
the longstanding issue of NuTeV anomaly could be solved.
The NuTeV anomaly indicates the weak-mixing-angle difference for
the NuTeV neutrino DIS measurement from other data average.
It is obtained in the NuTeV analysis by using 
the Paschos-Wolfenstein relation for neutral- and 
charged-current neutrino cross sections
\cite{nutev-anomaly}.
However, we need to take into account various small correction
factors to this relation:
$ R_A^- = (\sigma_{NC}^{\nu A} - \sigma_{NC}^{\bar\nu A})
        / (\sigma_{CC}^{\nu A} - \sigma_{CC}^{\bar\nu A}) 
  = 1/2 - \sin^2 \theta_W + O(\varepsilon_n) 
   + O(\varepsilon_i, \varepsilon_v, \varepsilon_{s_v}, \varepsilon_{c_v}) 
   + \cdots$.
Here, $\varepsilon_n$ is the neutron-excess correction in the iron nucleus
and it was taken into account in the NuTeV analysis. However, the other
corrections could be the source of the NuTeV anomaly.
The $\varepsilon_i$ is the isospin-violation effects on the PDFs,
$\varepsilon_v$ is the nuclear modification difference between
$u_v$ and $d_v$ in the iron, 
$\varepsilon_{s_v}$ ($\varepsilon_{c_v}$) is the valence strange (charm) 
quark ($q_v \equiv q - \bar q$) effect.
Of course, there is no net strangeness and charm in the nucleon
and ordinary nuclei, so that $\int dx s_v (x) = \int dx c_v (x)=0$ should
be satisfied, but it does not mean no $x$ dependence: $s_v (x) \ne 0$
and $c_v (x) \ne 0$.
All of these corrections are difficult to be determined at this stage;
however, studies of accurate PDFs from both theoretical and experimental
sides could lead to a finding on the source for the NuTeV anomaly.

There is recent progress on the PDFs in lattice QCD \cite{lattice-PDFs}.
The PDFs are defined by matrix elements of two-field correlators 
with gauge links between the fields to satisfy the color gauge 
invariance. Instead of lightcone-separated field correlators,
we may define quasi-PDFs with an equal-time separation, so that
it becomes possible to calculate the quasi-PDFs by lattice QCD.
Both PDFs agree in the infinite momentum limit $p_z \to \infty$.
This is a numerically challenging project for obtaining the PDFs 
at large $p_z$ in lattice QCD; however, it could be a promising 
future direction of theoretical PDFs from QCD.
In particular, if there is no sufficient experimental
information on some quantities, the lattice QCD studies will 
provide us a guideline.

As for the future prospect of the unpolarized PDFs of the nucleon,
we expect that the PDFs will be improved steadily with measurements 
of LHC and future experimental data of EIC and LHeC. 
Another interesting prospect is to study a real global analysis
with not only the unpolarized nucleonic data but also together with 
nuclear and polarized ones.
On the other hand, it is interesting to test the current PDFs
in ultra-high energy reactions which cannot be attained by man-made accelerators.
There exist ultra-high energy cosmic rays which arrive on the earth.
By extending our current knowledge of high-energy hadron reactions,
air-shower codes have been developed. The leading shower 
development is dominated by forward physics, namely described mainly
by the Regge and Pomeron theories. It is an interesting topic; however,
for a direct connection with the PDFs, we could test them by cross sections
of ultra-high energy neutrinos. The first IceCube neutrino cross sections
were recently reported up to the $10^{15}$ eV range \cite{icecube-2017}. 
Although the IceCube cross sections are slightly larger than 
the standard-model estimates, they are consistent within 
experimental uncertainties.

\section{Nuclear PDFs}
\label{nuclear-pdfs}

In 1970's, there was a prejudice that nuclear effects, which are
small energy-momentum scale phenomena of the order of 10-100 MeV, do not change
DIS structure functions measured in the range of 10-100 GeV.
Although nucleon Fermi-motion effects were noticed at the early stage, 
the EMC (European Muon Collaboration) discovery on
nuclear modification of $F_2$ was rather surprising.
It used to be considered as the first finding on an explicit quark
signature in nuclear physics. However, it is generally very difficult to 
pin down such an effect even though nuclei are dense systems of nucleons
with the average nucleon separation 2.2 fm, which is almost equal to
the nucleon diameter 1.8 fm.
It was found later that usual nuclear mechanisms,
in terms of nuclear binding and Fermi motion, could describe
the major part of the nuclear modifications of $F_2$ at $x>0.2$.
Therefore, although the EMC finding may contain important discoveries,
it is not easy to draw a solid conclusion on a new hadron-physics
mechanism.

In any case, the nuclear modifications of the PDFs should be understood
accurately for practical purposes, for example, 
in order to investigate properties of quark-gluon plasma 
in heavy-ion reactions as shown in 
Fig.\,\ref{fig:hard-cross}. In addition, neutrino oscillation measurements 
become accurate and it is the stage to probe CP violation
in the lepton sector. For this purpose, we need to understand
neutrino-nucleus cross sections within about 5\% accuracy,
for example, because the T2K target is water which contains 
the oxygen nucleus. The largest systematic error comes from
neutrino-nucleus interaction part, although the error is
significantly reduced by near-detector measurements.
T2K neutrino beam energies are not large enough
to be sensitive to the DIS process;
however, accurate nuclear PDFs (NPDFs) are important factors in
Fermilab neutrino experiments in the several GeV region.

\begin{wrapfigure}[12]{r}{0.45\textwidth}
    \hspace{-0.40cm}
\begin{minipage}[c]{0.45\textwidth}
    \vspace{-0.30cm}
   \begin{center}
    \includegraphics[width=6.5cm]{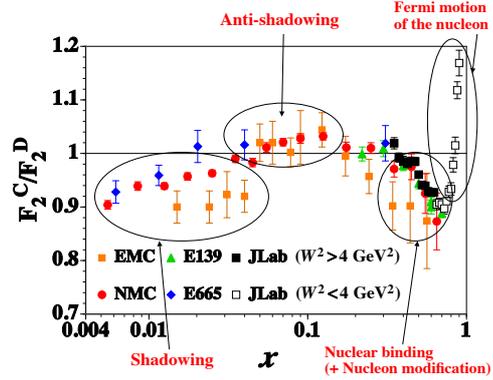}
   \end{center}
\vspace{-0.85cm}
\caption{Nuclear modifications in $F_2^C / F_2^D$.}
\label{fig:F2C-D}
\vspace{-0.5cm}
\end{minipage}
\end{wrapfigure}

For example, nuclear modifications are measured for the carbon nucleus
by the ratio $F_2^C / F_2^D$ in Fig.\,\ref{fig:F2C-D}.
In the small-$x$ region at $x<0.05$, there are negative modifications
typically 5-10\% for the carbon, and it is called shadowing.
The virtual photon could fluctuate into a $q\bar q$ (or a vector meson state)
at small $x$. The $q\bar q$ pair (or vector mesons) strongly interacts with
surface nucleons, so that the projectile does not interact with internal
nucleons, namely they are shadowed by the surface ones.
The $q\bar q$ propagation length is estimated as 
$\lambda = 1/|E_V -E_\gamma | = 2 \nu / (Q^2 + M_V^2 ) 
 = 0.2 \ \text{fm} /x >$2 fm at $x<0.1$. 
It becomes larger than the average separation of nucleons in a nucleus,
and then multiple scattering occurs. Using multiple scattering theory,
we can describe shadowing phenomena of $F_2$.
At medium $x$, there are also negative nuclear modifications, and the ratio
increases at large $x\, (>0.7)$ due to the nucleon Fermi motion. 
This part ($x>0.3$) is described by the convolution picture. 
Namely, the nuclear structure function
$F_2^A$ is calculated by the convolution of nucleon energy-momentum
distribution in a nucleus with the nucleon structure 
function $F_2^N$. A slight shift in the energy-momentum distribution
due to the binding energy results in the 5-10\% modifications
at medium $x$, and the Fermi motion modifies the function at large $x$.
The modification is positive at $x =0.1$ and it is called anti-shadowing
because of the opposite effect to the shadowing. It is necessary 
to have such positive effects to satisfy the baryon-number,
charge, and momentum conservations for a nucleus, but its physics
mechanism is not well studied so far.

There are two ways to determine the NPDFs. One is to obtain
nuclear modification functions as typically shown in Fig.\,\ref{fig:F2C-D},
and the other is to obtain the absolute NPDFs.
The advantage of the first one is that we could possibly
avoid obtaining unphysical NPDFs in the regions where there is 
no or few experimental data.
Furthermore, the $x$ dependence of the nuclear modifications is
roughly known and they are within 20-30\% even for large nuclei,
and functional forms of the modifications could be roughly the same 
for all the partons.
However, neutrino data are not provided in the ratio form
$e.g.$ with the deuteron, so that one has different treatments 
between the charged-lepton and neutrino data. 
The second method defines the absolute NPDFs on the same footing with
the nucleon ones, which is the advantage as investigated 
by the nCTEQ collaboration.

\vfill\eject 

\begin{wrapfigure}[16]{r}{0.38\textwidth}
    \hspace{-0.40cm}
\begin{minipage}[c]{0.38\textwidth}
    \vspace{0.20cm}
   \begin{center}
     \includegraphics[width=4.6cm]{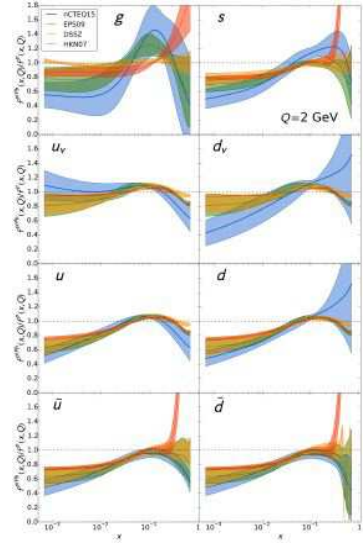}
   \end{center}
\vspace{-0.60cm}
\caption{PDF modifications in iron \cite{ncteq-2016}.}
\label{fig:npdf-pb}
\vspace{-0.60cm}
\end{minipage}
\end{wrapfigure}

The PDF modifications are shown for the nCTEQ analysis 
in Fig.\,\ref{fig:npdf-pb} at $Q^2 = 2^2$ GeV$^2$ 
in comparison with other distributions \cite{ncteq-2016}.
All the analysis results roughly agree with each other. 
The gluon modification is not well determined. Especially, the magnitude
of the gluon shadowing is not very clear at this stage, although
there could be a hint from $J/\psi$ production in the ultra-peripheral
heavy-ion collisions \cite{j/psi-upc}. As seen in Fig.\,\ref{fig:F2C-D},
there is no datum at $x<0.004$ for nuclear structure functions. However,
the situation should be changed significantly by the EIC project where
small-$x$ ($\sim 10^{-3}$) measurements will be done. Then, the scaling
violation could constraint the gluon shadowing accurately.
Furthermore, flavor dependent nuclear modifications could be measured
at JLab by the parity-violating DIS and 
at Fermilab by Drell-Yan processes with nuclear targets.

\section{Polarized PDFs}
\label{polarized-pdfs}

\subsection{Polarized PDFs of spin-1/2 nucleon}
\label{spin-1/2-nucleon}

\begin{wrapfigure}[22]{r}{0.38\textwidth}
   \vspace{-0.7cm} 
   \begin{center}
     \includegraphics[width=5.7cm]{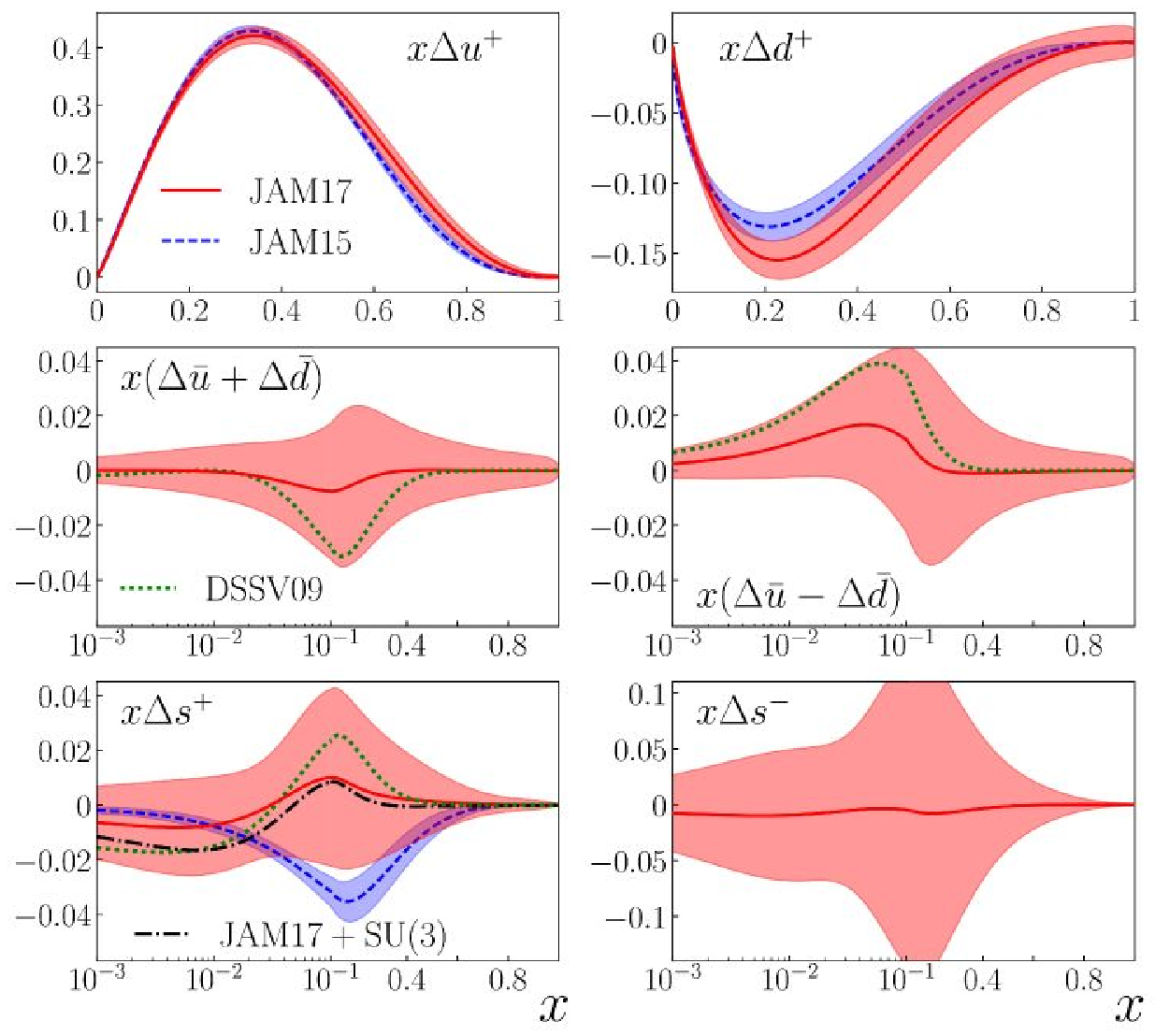}
\vspace{-0.80cm}
\caption{Polarized PDFs of nucleon \cite{jam-2017}.}
\label{fig:jam-2017}
\vspace{+0.60cm}
     \includegraphics[width=4.5cm]{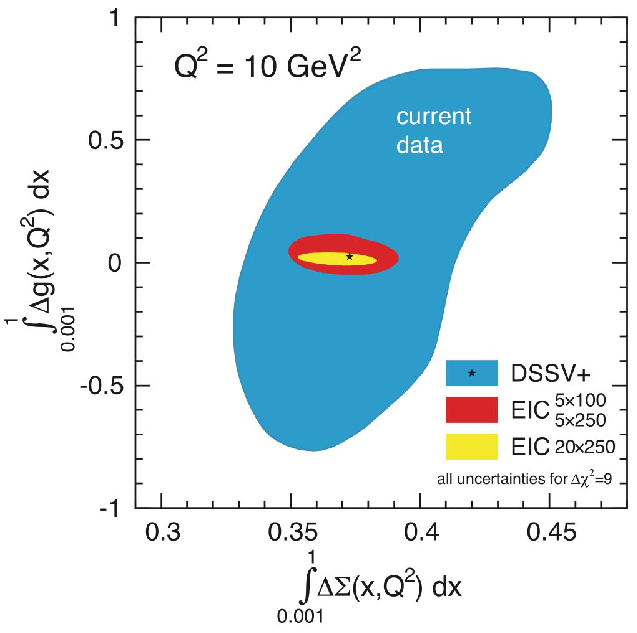}
\vspace{-0.40cm}
\caption{Gluon polarization at EIC \cite{eic-2016}.}
\label{fig:gluon-pol}
\end{center}
\end{wrapfigure}

The longitudinally-polarized PDFs have been investigated 
since 1980's, and we now know the quark and antiquark distributions
from polarized DIS processes and polarized proton-proton collisions,
although flavor dependent antiquark and 
gluon distributions still have large errors. 
It is known that the determined polarized PDFs could 
be changed depending on fragmentation functions if the semi-inclusive DIS
data are included. In a recent JAM collaboration analysis,
fragmentation function data are included into their analysis 
data set toward a ``real'' global analysis \cite{jam-2017}.
The determined polarized PDFs are shown in Fig.\,\ref{fig:jam-2017}
with their uncertainties. The $\Delta u^+$ and $\Delta d^+$ distributions
are well determined, but the antiquark ones have large errors at this stage.

The gluon polarization has not been determined accurately.
There was a report in 2014 that a significant fraction
($\sim\ $40\%) of nucleon spin could be carried by the gluon
from an analysis with $\pi^0$-production measurements 
in the polarized pp collisions at RHIC. However, its uncertainty
is still large as shown in Fig.\,\ref{fig:gluon-pol} because
the small-$x$ region has not been probed in any experiments so far.
One of the major projects of EIC is to determine the gluon-spin
contribution to the nucleon spin precisely \cite{eic-2016}.
In particular, scaling-violation measurements will impose
a strong constraint for the gluon polarization, and 
it should be determined accurately by the future EIC data.

\begin{wrapfigure}[10]{r}{0.41\textwidth}
   \vspace{-0.7cm} 
   \begin{center}
     \includegraphics[width=4.2cm]{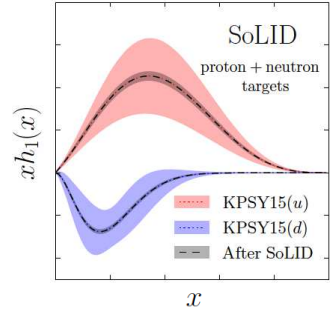}
\vspace{-0.40cm}
\caption{Transversity and SoLID \cite{trans-solid-2017}.}
\label{fig:trans-solid}
\end{center}
\end{wrapfigure}

These are longitudinally-polarized PDFs, whereas there are
transverse ones called transversity distributions. 
The quark transversity distributions have chiral-odd nature,
and they are decoupled from the gluon one in the $Q^2$ evolution.
Because transverse spin phenomena should be the same as
the longitudinal one at low energies, their difference
from the longitudinally-polarized PDFs should deepen
our knowledge on nucleon spin. It is interesting that 
their $Q^2$ evolution difference from the longitudinally-polarized PDFs 
probes a perturbative aspect of nucleon spin physics.
The transversity distributions are determined precisely by the future
SoLID (Solenoidal Large Intensity Device) experiment
as shown in Fig.\,\ref{fig:trans-solid} \cite{trans-solid-2017}.

\subsection{Polarized PDFs of spin-1 deuteron}
\label{spin-1-deuteron}

\begin{wrapfigure}[20]{r}{0.41\textwidth}
   \vspace{-0.7cm}
   \begin{center}
     \includegraphics[width=5.5cm]{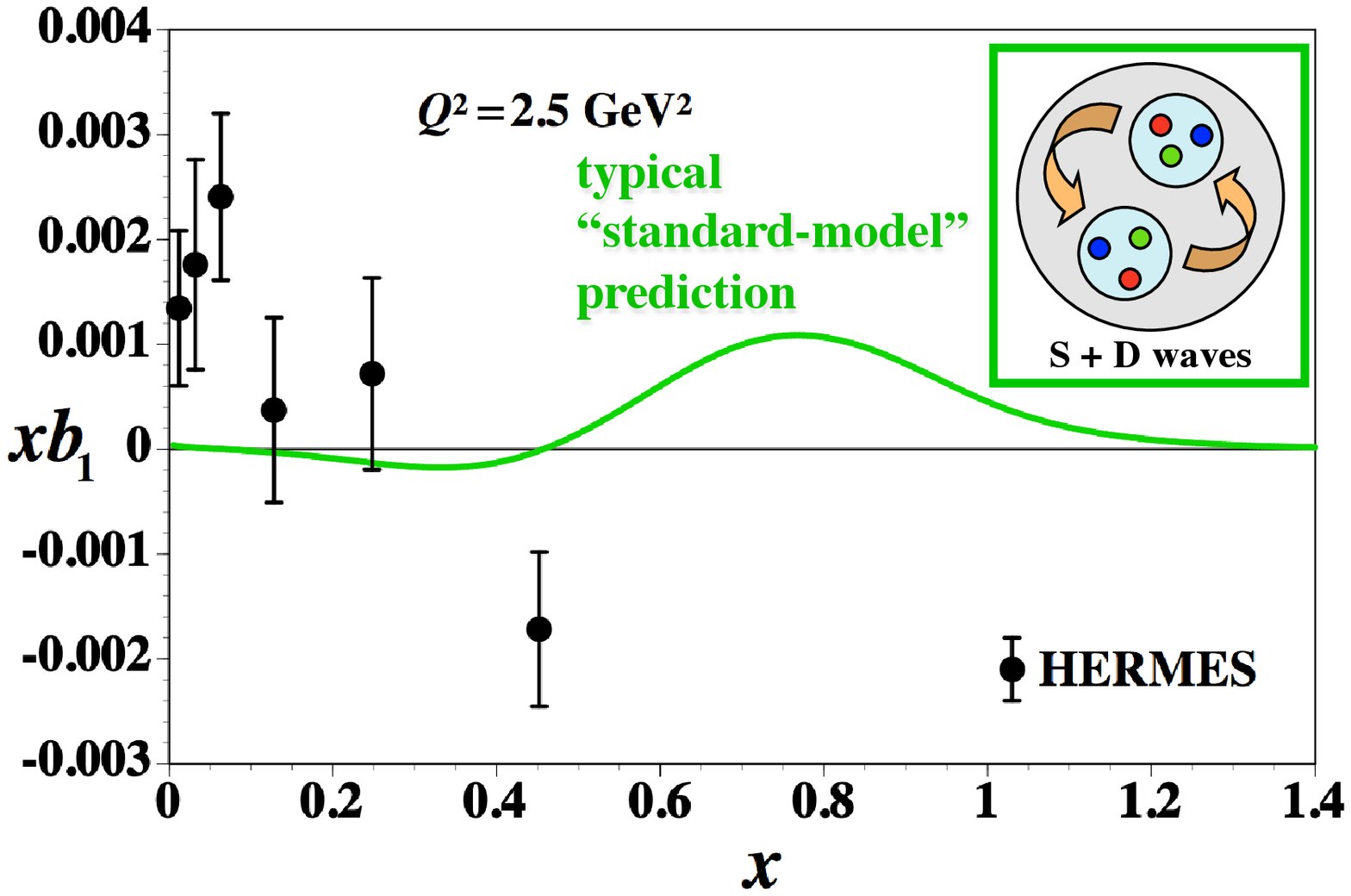}
\vspace{-0.40cm}
\caption{Standard convolution model for $b_1$ 
and HERMES data.}
\label{fig:b1-conv}
\vspace{+0.60cm}
     \includegraphics[width=5.5cm]{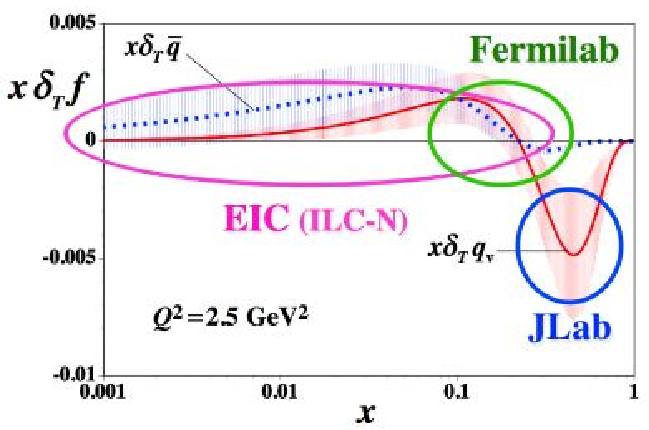}
\vspace{-0.40cm}
\caption{Future prospects on tensor-polarized PDFs.}
\label{fig:future-tensor}
   \end{center}
\end{wrapfigure}

It is known that the spin-1 deuteron has additional structure functions
to the ones in the spin-1/2 nucleon. The twist-2 ones are called
$b_1$ and $b_2$, which are related with each other by the Callan-Gross
type relation in the Bjorken scaling limit. The $b_1$ project will
start soon at JLab \cite{jlab-b1-fermilab}, whereas
there are HERMES measurements as shown in Fig.\,\ref{fig:b1-conv}. 

The deuteron structure has been investigated
for a long time at low energies and it is described as a bound system
of proton and neutron in mainly S wave with small D-wave admixture.
Using this standard model with a convolution formalism for the deuteron
structure function $b_1$, we obtained the solid curve \cite{b1-conv}, which is very
different from the HERMES measurements, as shown in Fig.\,\ref{fig:b1-conv}.
Since the HERMES errors are large, we had better wait for the JLab measurement. 
However, the large difference could suggest that a new hadron-physics
mechanism may be needed for explaining $b_1$ beyond the usual
standard convolution model. A new interesting field of high-energy 
spin physics could be created by future $b_1$ measurements at JLab
and EIC. In the parton model, $b_1$ is expressed by the tensor-polarized
PDFs, which could be also measured at Fermilab by proton-deuteron
Drell-Yan experiment E1039 \cite{jlab-b1-fermilab}.
The kinematical regions of these future projects are shown together
with possible tensor-polarized PDFs to explain the HERMES data
\cite{sk-tensor} in Fig.\,\ref{fig:future-tensor}. 

\vfill\eject
\section{Fragmentation functions}
\label{ffs}

Fragmentation functions (FFs) indicate probabilities of parton fragmentation
into hadrons, and they are used for calculating high-energy hadron 
reaction cross sections in Fig.\,\ref{fig:hard-cross}.
They are determined mainly by hadron-production measurements in electron-positron
annihilation. The Belle and BaBar collaborations published accurate data 
on the hadron productions at the center-of-mass energy
of about 10 GeV. In comparison with previous measurements at SLD and LEP
at the Z mass, the gluon fragmentation function is determined more
accurately through the scaling violation \cite{HKKS-2016}.

\begin{wrapfigure}[12]{r}{0.55\textwidth}
   \vspace{-0.6cm} 
   \begin{center}
     \includegraphics[width=7.5cm]{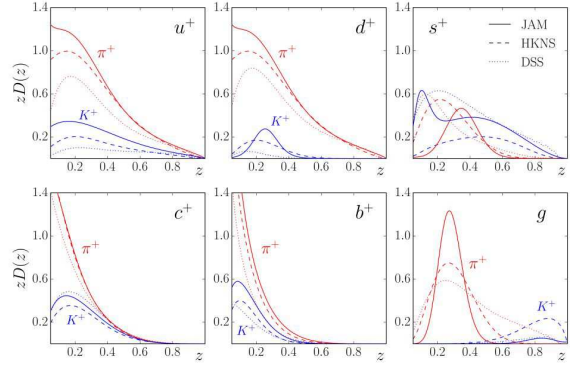}
   \end{center}
\vspace{-0.80cm}
\caption{Determination of fragmentation functions \cite{JAM-FFs-2016}.}
\label{fig:JAM-FFs-2016}
\vspace{-0.60cm}
\end{wrapfigure}

In Fig.\,\ref{fig:JAM-FFs-2016}, typical global analysis results 
\cite{JAM-FFs-2016} are shown for the pion and kaon FFs. 
These functions are obtained by analyzing $e^+ e^-$ data,
and they are compared with previous analysis results of HKNS and DSS.
There are variations among the groups, but they are roughly consistent
with each other within the uncertainties.
We should note that a variation in strange fragmentation
function could result in a significant change in the unpolarized
and polarized strange-quark distributions in the nucleon.
Therefore, for a precise determination of nucleonic PDFs in future,
it is necessary to have accurate FFs because hadron-production
data are included in many global analysis of PDFs.

\section{Hadron tomography by three-dimensional structure functions}
\label{3d-sfs}

\begin{wrapfigure}[10]{r}{0.55\textwidth}
   \vspace{-0.7cm} 
   \begin{center}
     \includegraphics[width=8.3cm]{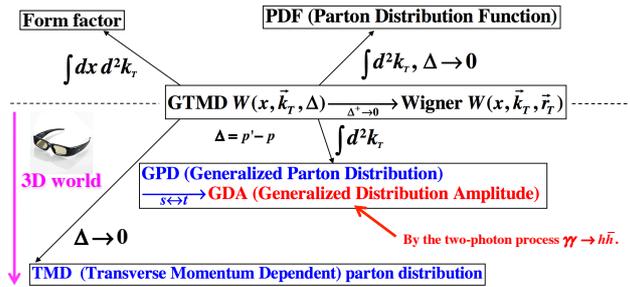}
   \end{center}
\vspace{-0.80cm}
\caption{Wigner distribution and 3D structure functions.}
\label{fig:wigner-3d}
\vspace{-0.60cm}
\end{wrapfigure}

For clarifying internal structure of the nucleon, we have been
investigating elastic form factors and the PDFs.
In order to find the origin of the nucleon spin, orbital angular
momenta of quarks and gluons become important factors and they could be
probed by three-dimensional structure functions called
generalized parton distributions (GPDs).
The $s$-$t$ crossed quantities of the GPDs are 
generalized distribution amplitudes (GDAs).
There are other popular 3D functions called
transverse-momentum-dependent parton distributions (TMDs).
The form factors, PDFs, GPDs, GDAs, and TMDs are obtained 
from the generating function, the Wigner distribution
or generalized TMDs as shown in Fig.\,\ref{fig:wigner-3d}.

\vfill\eject
\subsection{Transverse-momentum-dependent parton distributions (TMDs)}
\label{tmds}

\begin{wrapfigure}[17]{r}{0.38\textwidth}
   \vspace{-0.8cm} 
   \begin{center}
     \includegraphics[width=4.5cm]{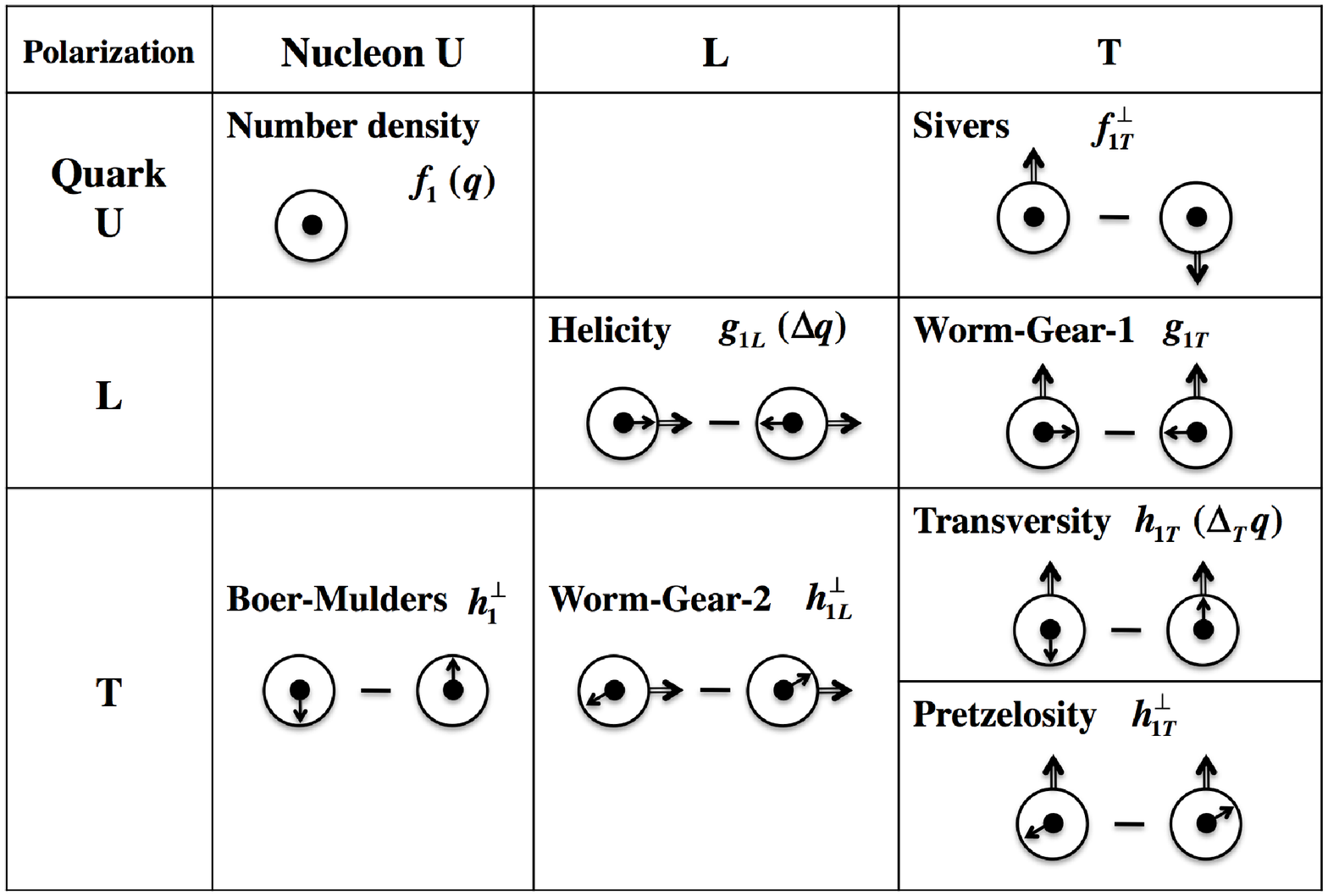}
\vspace{-0.20cm}
\caption{Various TMDs with nucleon and quark polarizations.}
\label{fig:tmd-1}
\vspace{+0.30cm}
     \includegraphics[width=5.0cm]{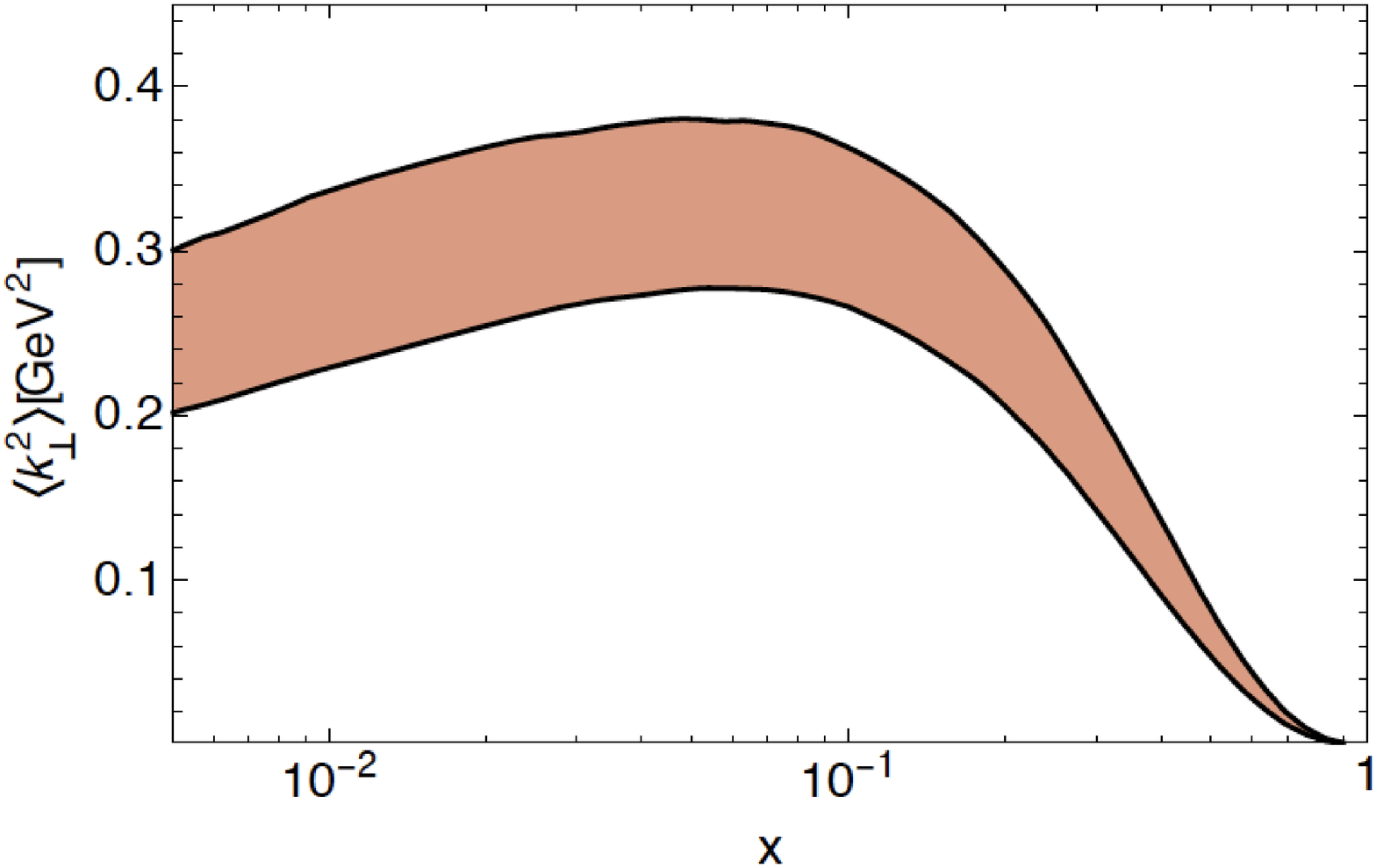}
\vspace{-0.40cm}
\caption{Transverse-momentum squared average \cite{unpol-tmds-2017}.}
\label{fig:kT2}
   \end{center}
\end{wrapfigure}

In Fig.\,\ref{fig:tmd-1}, the TMD types are shown by indicating 
polarizations of nucleon and quark. 
We have already explained the unpolarized PDFs, 
the longitudinally-polarized PDFs, 
and the transversity distributions.
There are corresponding TMDs $f_1$, $g_{1L}$, and $h_{1T}$.
In addition, there are Boer-Mulders ($h_1^\perp$), 
Sivers ($f_{1T}^\perp$), worm-gear ($g_{1T}$, $h_{1L}^\perp$), 
pretzelosity ($h_{1T}^\perp$) distributions depending on the polarizations.

The unpolarized TMDs are obtained in Ref.\,\cite{unpol-tmds-2017},
where the Gaussian function is assumed for the transverse distributions.
Using semi-inclusive DIS, Drell-Yan, and Z-boson production data,
they determined the TMDs by restricting the analysis to 
the low-transverse-momentum region.
A typical average transverse-momentum squared 
$\langle k_\perp^{\, 2} \rangle$
is shown as the function of $x$ in Fig.\,\ref{fig:kT2}.
At large $x$ where valence quarks dominate,
the average $k_T^2$ is small, and it becomes larger at small $x$
in the sea-quark region, and then it stays roughly the same at $x<0.1$.

\begin{wrapfigure}[9]{r}{0.38\textwidth}
   \vspace{-0.7cm} 
   \begin{center}
     \includegraphics[width=4.8cm]{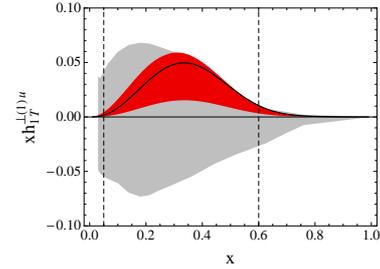}
   \end{center}
\vspace{-0.70cm}
\caption{Pretzelosity and SoLId \cite{Pretzelosity-solid}.}
\label{fig:Pretzelosity-solid}
\vspace{-0.60cm}
\end{wrapfigure}

Precision studies of the TMDs will be done by the JLab SoLID project.
There are studies on determinations of the Sivers \cite{tmd-sivers}, 
Pretzelosity \cite{tmd-pretz}, and Boer-Mulders \cite{tmd-bm}.
However, the current distributions have large uncertainties in general, 
and the SoLID experiment will improve the situation drastically 
in the near future by accurate measurements. It is typically shown in 
Fig.\,\ref{fig:Pretzelosity-solid}, where the $u$-quark
pretzelosity distribution is shown. Currently, the uncertainty
is too large and we are not sure whether it is even positive or negative
distribution. The SoLID project will certainly clarify the distribution
with small uncertainties \cite{Pretzelosity-solid}. 
Furthermore, the EIC project probe the TMDs in a smaller-$x$ region. 
By these future experimental measurements, the TMD physics will be
developed significantly in 2020's.

\subsection{Origin of nucleon spin by 3D structure functions}
\label{n-spin}

\begin{wrapfigure}[8]{r}{0.32\textwidth}
   \vspace{-0.6cm} 
   \begin{center}
     \includegraphics[width=3.2cm]{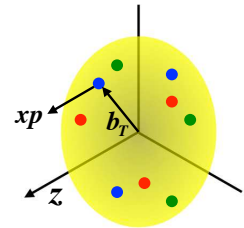}
\vspace{-0.20cm}
\caption{3D view of hadron.}
\label{fig:3D-hadron}
   \end{center}
\end{wrapfigure}

We explained the situation of the origin of the nucleon spin, namely
the longitudinally-polarized PDFs, in Sec.\ref{polarized-pdfs}.
There is still large uncertainty in the gluon contribution. 
However, the remaining possibility is the effect of
partonic orbital angular momenta. It can be determined experimentally
by the GPDs which are measured in deeply virtual Compton scattering (DVCS).
For example, the second moments of quark GPDs, 
$H_q (x,\xi,t=0)$ and $E_q (x,\xi,t=0)$,
are related to the orbital-angular-momentum contribution $L_q$ as
\begin{align}
J_q & = \frac{1}{2} \int_{-1}^1 dx x [ H_q (x,\xi,t=0) + E_q (x,\xi,t=0) ]
\nonumber \\
 & = \frac{1}{2} \Delta q + L_q ,
\vspace{-0.20cm}
\nonumber
\label{eqn:oam}
\end{align}

\begin{wrapfigure}[11]{r}{0.46\textwidth}
   \vspace{-0.4cm} 
   \begin{center}
     \includegraphics[width=7.0cm]{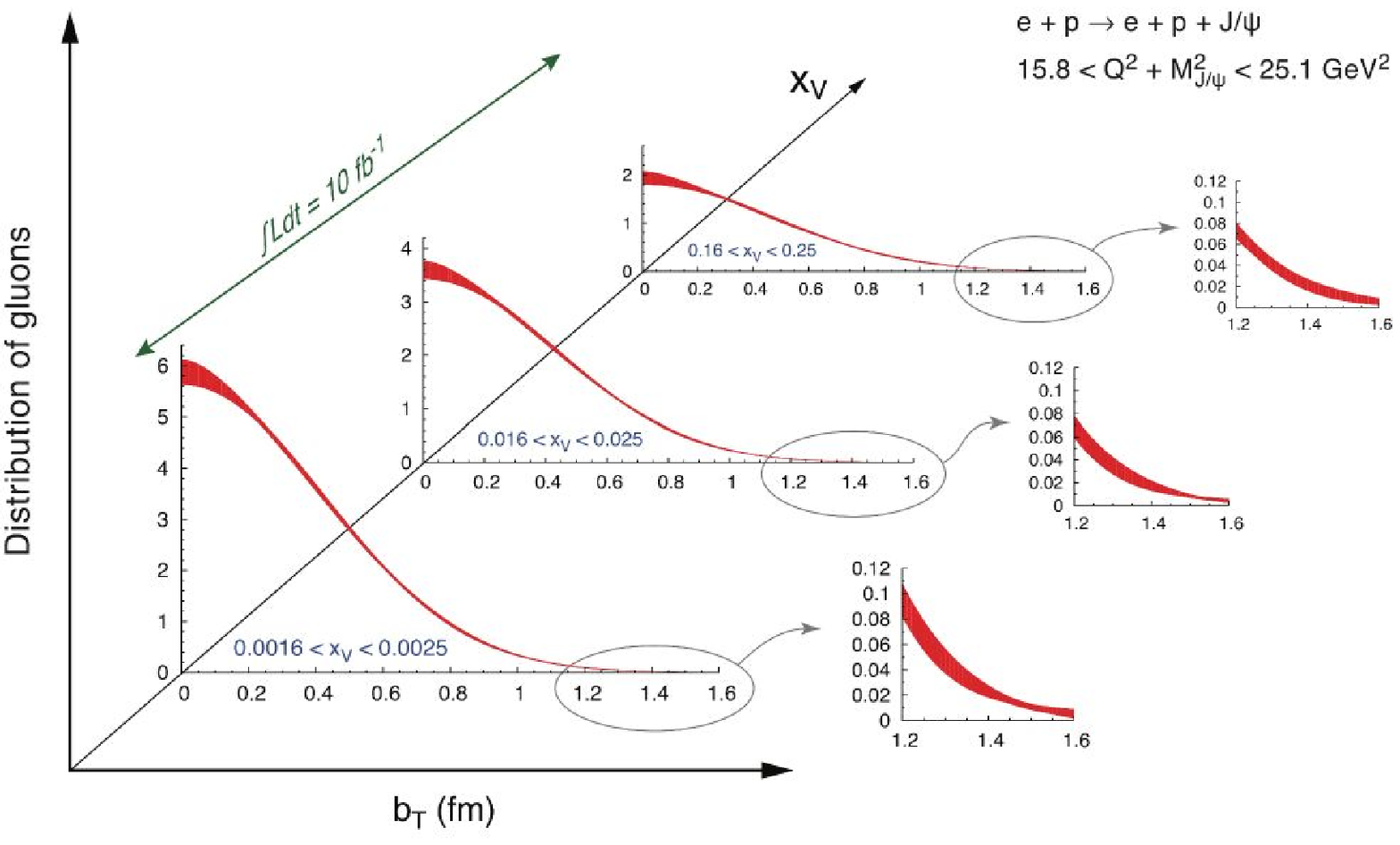}
\vspace{-0.80cm}
\caption{GPD studies at EIC \cite{eic-2016}.}
\label{fig:gpds-eic}
   \end{center}
\end{wrapfigure}

\noindent
by finding the quark-spin 
contribution $\Delta q$ from other experiments. 
The Fourier transforms of the GPDs in the transverse coordinates
in Fig.\,\ref{fig:3D-hadron}
indicate partonic transverse distributions as shown
in Fig.\,\ref{fig:gpds-eic}.

Because the GPDs have three variables, it is difficult to determine
accurately in the whole kinematical region. The DVCS cross section
is given by the GPDs integrated over $x$, so that we need assumptions
on their functional forms on $x$, $\xi$, and $t$, or by the GPDs
with the constraint $x=\xi$ or $-\xi$.
The GPDs are measured by the virtual Compton scattering,
and measurements are in progress in the JLab and COMPASS experiments.
The current GPDs are not well constrained \cite{gpds}, 
so that it is one of major purposes of the future EIC project 
to determine them accurately.
Furthermore, there are possibilities to measure the GPDs
at hadron facilities such as J-PARC, for example, by using
exclusive Drell-Yan $\pi^- + p \to \mu^+ \mu^- + n$
and exclusive hadron reactions, $e.g.$ $N+N \to N + \pi + B$
\cite{j-parc-gpds}
at the high-momentum beamline under construction right now.
In future, hadron facilities could provide an alternative
way to probe 3D structure of hadrons.

\begin{wrapfigure}[9]{r}{0.53\textwidth}
    \hspace{-0.40cm}
\begin{minipage}[c]{0.50\textwidth}
    \vspace{-0.20cm}
   \begin{center}
     \includegraphics[width=8.3cm]{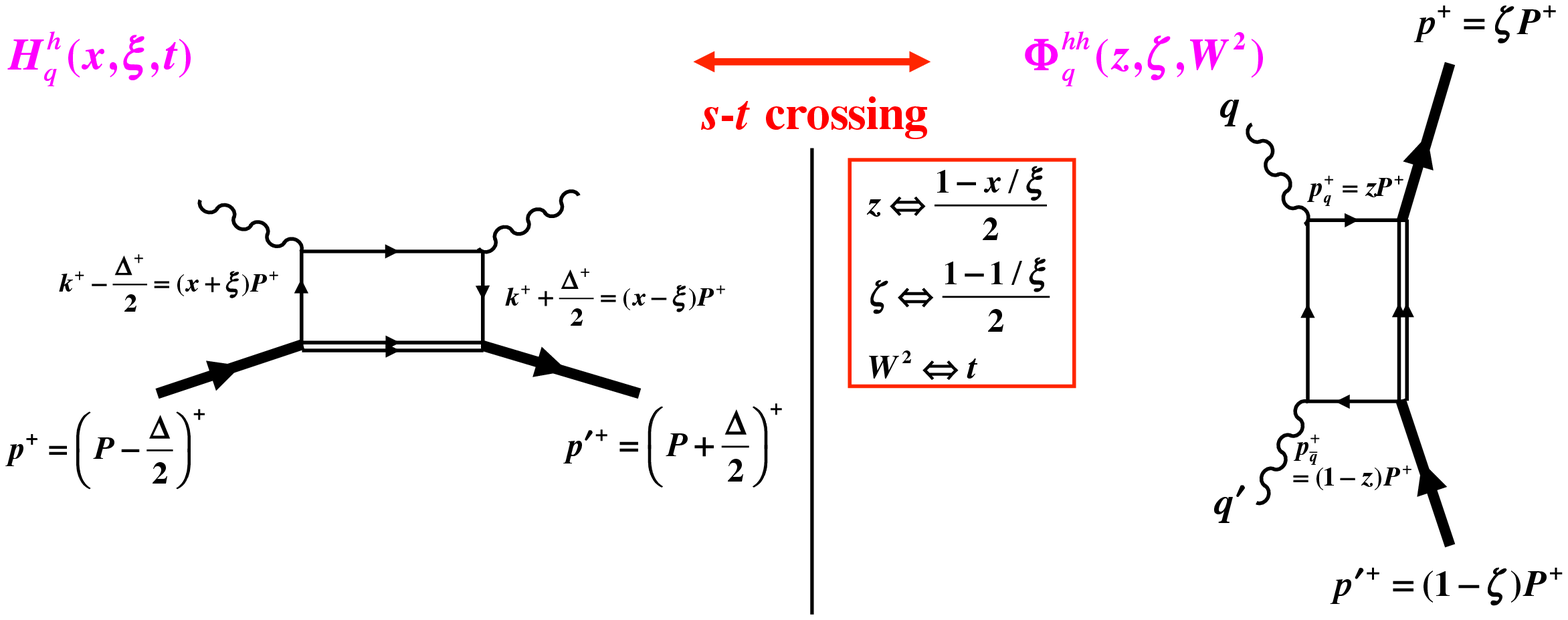}
   \end{center}
\vspace{-0.60cm}
\caption{GPDs and GDAs.}
\label{fig:3D-gpd-gda}
\vspace{-0.60cm}
\end{minipage}
\end{wrapfigure}

The GPDs are determined by the DVCS $\gamma^* h \to \gamma h$
on the left-hand side of Fig.\,\ref{fig:3D-gpd-gda}, 
and there are $s$-$t$ crossed quantities,
the GDAs, as shown on the right-hand side \cite{kk-2014,kst-2018}.
The GDAs are measured in the two-photon process 
$\gamma^* \gamma \to h \bar h$, which is possible
at electron-positron collider such as KEKB and future ILC.
The GPDs and GDAs are connected by the $s$-$t$ crossing, so that
both quantities are valuable for probing 3D structure of hadrons
and origin of the nucleon spin.

\subsection{Tomography of exotic hadron candidates}
\label{exotics}

In the last decade, there have been reports on a number of 
exotic-hadron candidates. However, it is not obvious whether
the findings indicate ``real'' exotic hadrons only by
global observables such as masses, spins, decay widths, $etc$.
Furthermore, because the quark number is not a conserved quantity,
we need to consider a possible way to find exotic configurations. 
In this respect, high-energy hadron reactions could be useful tools to find 
the internal structure. For example, there is a method to
find an internal constituent number involved in 
a hard exclusive reaction. It is called the constituent-counting rule
in perturbative QCD \cite{counting}.

\vfill\eject

\begin{wrapfigure}[11]{r}{0.40\textwidth}
    \hspace{-0.30cm}
\begin{minipage}[c]{0.40\textwidth}
    \vspace{0.10cm}
   \begin{center}
     \includegraphics[width=5.6cm]{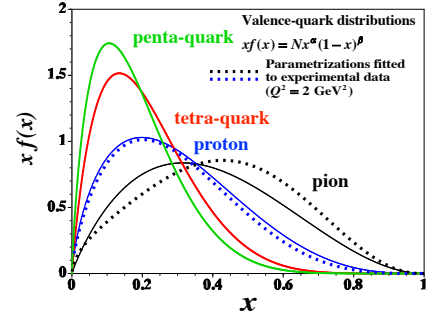}
   \end{center}
\vspace{-0.75cm}
\caption{Valence-quark distributions of exotic hadrons \cite{kk-2014}.}
\label{fig:exotic-pdfs}
\vspace{-0.60cm}
\end{minipage}
\end{wrapfigure}

On the other hand, there are possibilities to use the 3D structure
functions which we have been discussing in this section \cite{kk-2014}.
A simple functional form of the GPDs could be written as
$H_q^h (x,t) = q(x) F(t,x)$, where $F(t,x)$ is the transverse
form factor at $x$. We can predict the valence-quark distributions
of exotic hadrons in Fig.\,\ref{fig:exotic-pdfs}
by assuming the functional form $x f(x) = N x^\alpha (1-x)^\beta$ 
with the constraint of valence-quark number, 
counting rule of perturbative QCD,
and momentum fraction carried by valence quarks.
Because momenta carried by more quarks (4 or 5)
in the tetra- and penta-quark hadrons, the distributions
shift to the smaller-$x$ region. The transverse spread should be
also different depending on a compact quark-bound state or 
a diffusive molecular state, which could be observed
in the transverse form factor of the GPDs.
Of course, there is no stale fixed target for unstable exotic
hadrons, so that such GPDs studies are not directly possible
except for the transition GPDs.
However, the exotic-hadron tomography should be possible
by using the GDAs \cite{kk-2014}, which may be called timelike GPDs, 
because the exotic-hadron-pair productions are possible as shown
in Fig.\,\ref{fig:3D-gpd-gda}.
Because exotic-hadron cross sections are small in general,
it is not easy but it could be an interesting future project.

\subsection{Origin of gravitational sources and hadron masses}
\label{mass}

\begin{wrapfigure}[12]{r}{0.55\textwidth}
    \hspace{-0.40cm}
\begin{minipage}[c]{0.50\textwidth}
    \vspace{-0.30cm}
   \begin{center}
    \includegraphics[width=8.6cm]{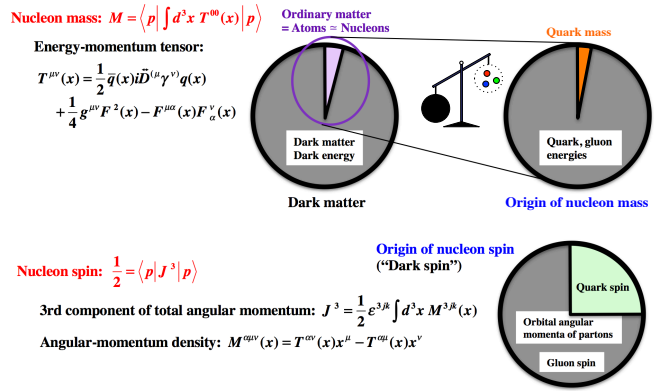}
   \end{center}
\vspace{-0.80cm}
\caption{Origins of nucleon mass and spin.}
\label{fig:hadron-mass-spin}
\vspace{-0.4cm}   
\end{minipage}
\end{wrapfigure}

We discussed the studies on the origin of nucleon spin 
in polarized DIS and DVCS. The nucleon spin is defined by
the matrix element of the angular-momentum tensor expressed by
the energy-momentum tensor, whereas the nucleon mass
is defined by the matrix element of the energy-momentum tensor
as shown in Fig.\,\ref{fig:hadron-mass-spin}.
Therefore, studies to find the origin of nucleon spin 
are analogous to the ones of the origin of nucleon mass.
For the last twenty years, there have been discussions
on the gauge-invariant decomposition of nucleon spin,
and such studies settled down recently. The origin of nucleon
mass has been also discussed, but they are mainly by 
effective hadron models. However, time has come to
clarify it in terms of QCD with experimental confirmations.

As a part of such studies, we can investigate gravitational
form factors which indicate mass, pressure, and shear-force
distributions in hadrons in terms of quark and gluon degrees of freedom. 
Gravitational interactions are too weak to be studied directly 
in scattering experiments, so that they had been considered as
a purely theoretical topic. However, there is a way to probe
the gravitational form factors by the GPDs and GDAs.
The GPDs and GDAs are defined by the same non-local vector operator,
and their moments are given as
\begin{equation}
2 \left ( \frac{P^+}{2}  \right ) ^2
 \! \! \! \int_0^1 \! \! dz \, (2z-1)^{n-1} 
  \! \!  \int\frac{d y^-}{2\pi}e^{i (2z-1) P^+ y^- /2}
  \overline{q}(-y/2) \gamma^+ q(y/2) \Big |_{y^+ = \vec y_\perp =0}
\! \! \!
 = \overline q (0) \gamma^+ \!
 \left ( i \overleftrightarrow \partial^+  \right )^{n-1} 
\! \! \! \!
 q(0) .
\nonumber
\label{eqn:tensor-int}
\end{equation}

\begin{wrapfigure}[18]{r}{0.40\textwidth}
   \vspace{-0.7cm} 
   \begin{center}
     \includegraphics[width=5.0cm]{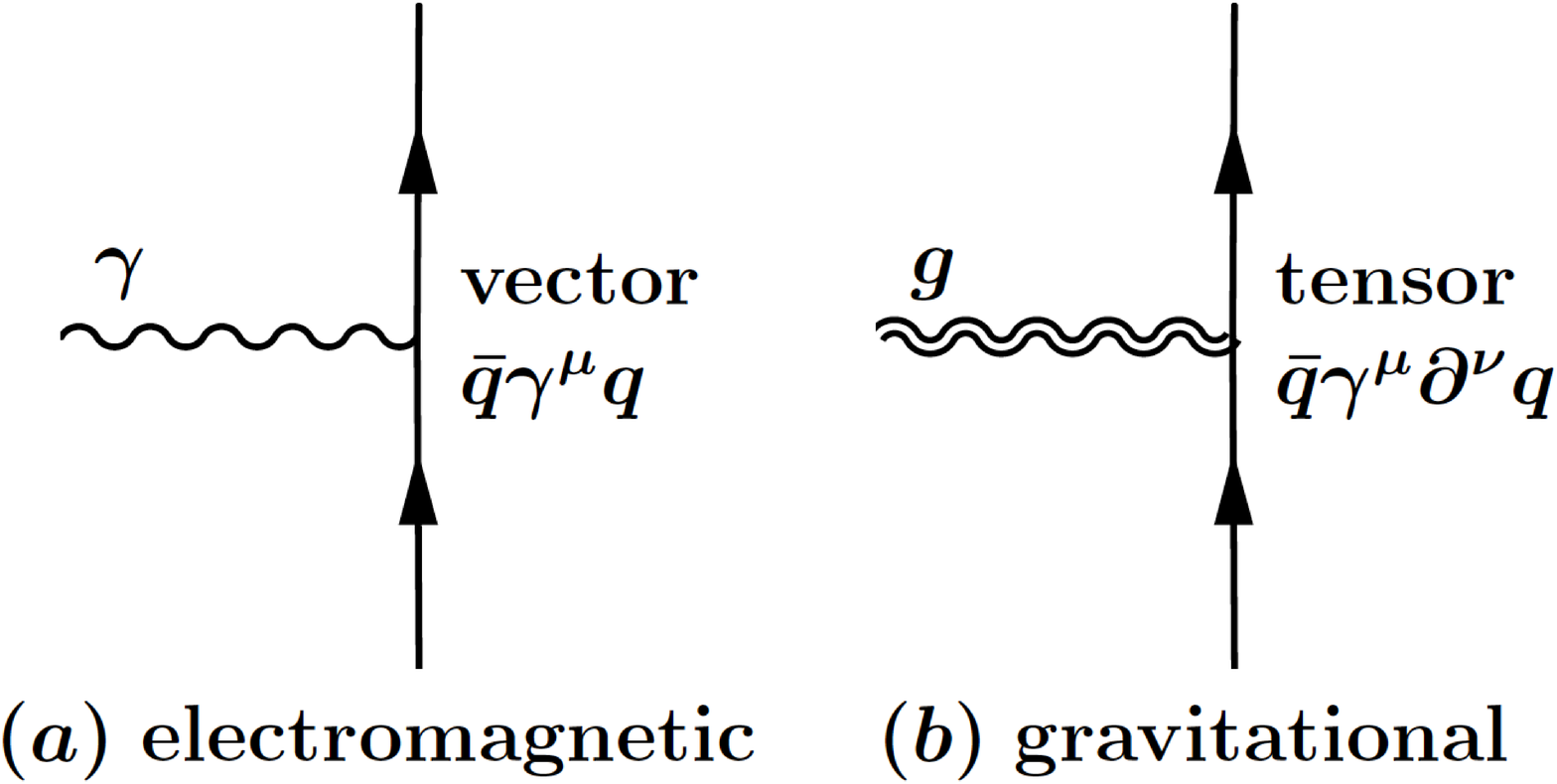}
   \end{center}
\vspace{-0.60cm}
\caption{Electromagnetic and gravitational interactions.}
\label{fig:elemag-grav}
   \vspace{+0.1cm}
   \begin{center}
     \includegraphics[width=4.8cm]{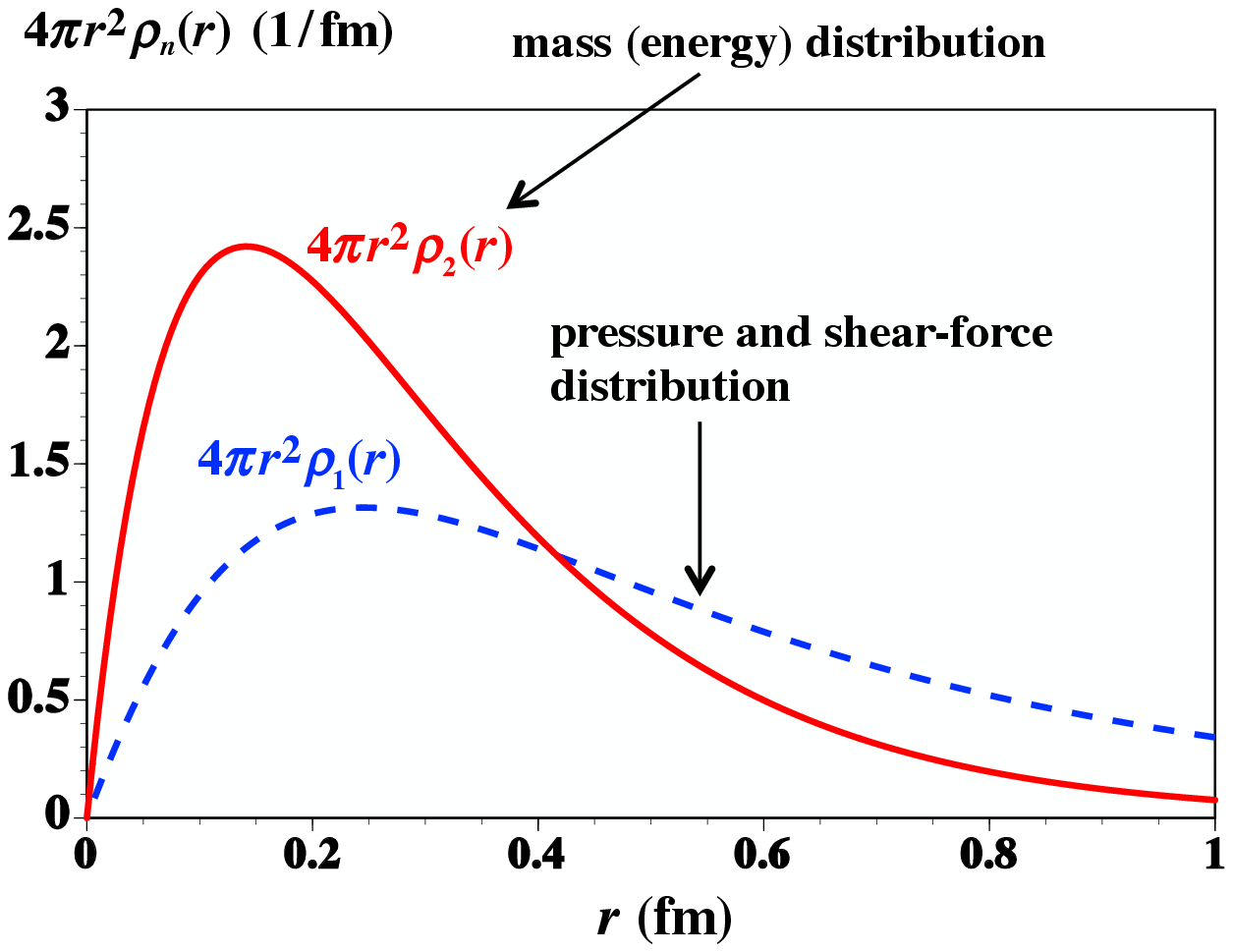}
   \end{center}
\vspace{-0.70cm}
\caption{Mass and mechanical (pressure /shear-force) distributions in pion.}
\label{fig:rho_pi}
\vspace{-0.60cm}
\end{wrapfigure}

\noindent
For $n=2$, this operator is the energy-momentum tensor
of a quark, and it is a source of gravity, whereas
it is the vector-type electromagnetic current for $n=1$,
as shown in Fig.\,\ref{fig:elemag-grav} \cite{kst-2018}.

Then, using KEKB experimental data, for example, for the two-photon process
$\gamma^* \gamma \to \pi^0 \pi^0$, we determined timelike gravitational
form factors, which are then converted to spacelike form factors
by using a dispersion relation.
They indicate mass (energy) distribution
as well as the pressure and shear-force distributions in the pion
as shown in Fig.\,\ref{fig:rho_pi},
and they are obtained from two gravitational form factors
$\Theta_1$ and $\Theta_2$.
From these form factors, we calculated the mass (energy) radius 
and the mechanical (pressure and shear force) radius
from $\Theta_2$ and $\Theta_1$, respectively.
They are calculated as
$\sqrt {\langle r^2 \rangle _{\text{mass}}} =0.56 \sim 0.69$ fm,
whereas the mechanical radius obtained from $\Theta_1$ is larger 
$\sqrt {\langle r^2 \rangle _{\text{mech}}} =1.45 \sim 1.56$ fm
\cite{kst-2018}.
This is the first report on the gravitational radii of a hadron
from actual experimental measurements.
It is interesting to find the possibility 
that the gravitational mass and mechanical radii 
could be different from the experimental charge radius 
$\sqrt {\langle r^2 \rangle _{\text{charge}}} =0.672 \pm 0.008$ fm
for the charged pion.
There are also recent studies on the gravitational form factors
\cite{mechanical}.
The KEKB was just upgraded, so that much accurate measurements will
be obtained in the near future for the GDAs.
We expect that the GDA studies will become popular as the GPD ones.
Similar studies should be possible by the GPDs, especially
if accurate DVCS measurements will be done at JLab and EIC.

\subsection{Nuclear composition of ultra-high-energy cosmic rays}
\label{cosmic-ray}

Ultra-high-energy cosmic ray physics is interesting not only
for studying the cosmic microwave background, especially for
confirming the GZK (Greisen-Zatsepin-Kuzmin) cutoff,
but also for hadron physics on small-$x$ and forward physics.
Another important connection, especially to the hadron tomography, is 
the determination of nuclear composition of ultra-high-energy 
cosmic rays. The nuclear composition, namely whether the cosmic
ray is proton or heavier nuclei such as iron, is determined
by the slant depth of shower maximum $X_{\text{max}}$
in the simulation of air shower by taking into account
cosmic-ray measurements on the earth surface. There are experimental 
indications that the ultra-high-energy cosmic rays are 
light nuclei according to recent Auger results
in the energy region $10^{17}$-$10^{20}$ eV.
Although typical simulation codes, EPOS, SIBYLL, QGSJET, and DPMJET
indicate similar energy dependencies, the situation could change
by the hadron-tomography studies. It is typically shown
in Ref.\,\cite{pk-2010} that the energy dependence could change
depending on the transverse structure of the nucleon.
Currently, studies on the 3D structure functions GPDs, GDAs, and TMDs
are in progress as explained in this article. We expect that
their relations with the cosmic-ray phenomena should be investigated
further in the near future, as well as the studies of
small/large-$x$ physics and Regge/Pomeron physics.

\subsection{Color entanglement}
\label{entanglement}

\begin{figure}[b!]
\vspace{0.10cm}
\begin{minipage}{\textwidth}
\begin{tabular}{lc}
\hspace{-0.30cm}
\begin{minipage}[c]{0.50\textwidth}
   \vspace{-0.0cm}
   \begin{center}
     \includegraphics[width=7.0cm]{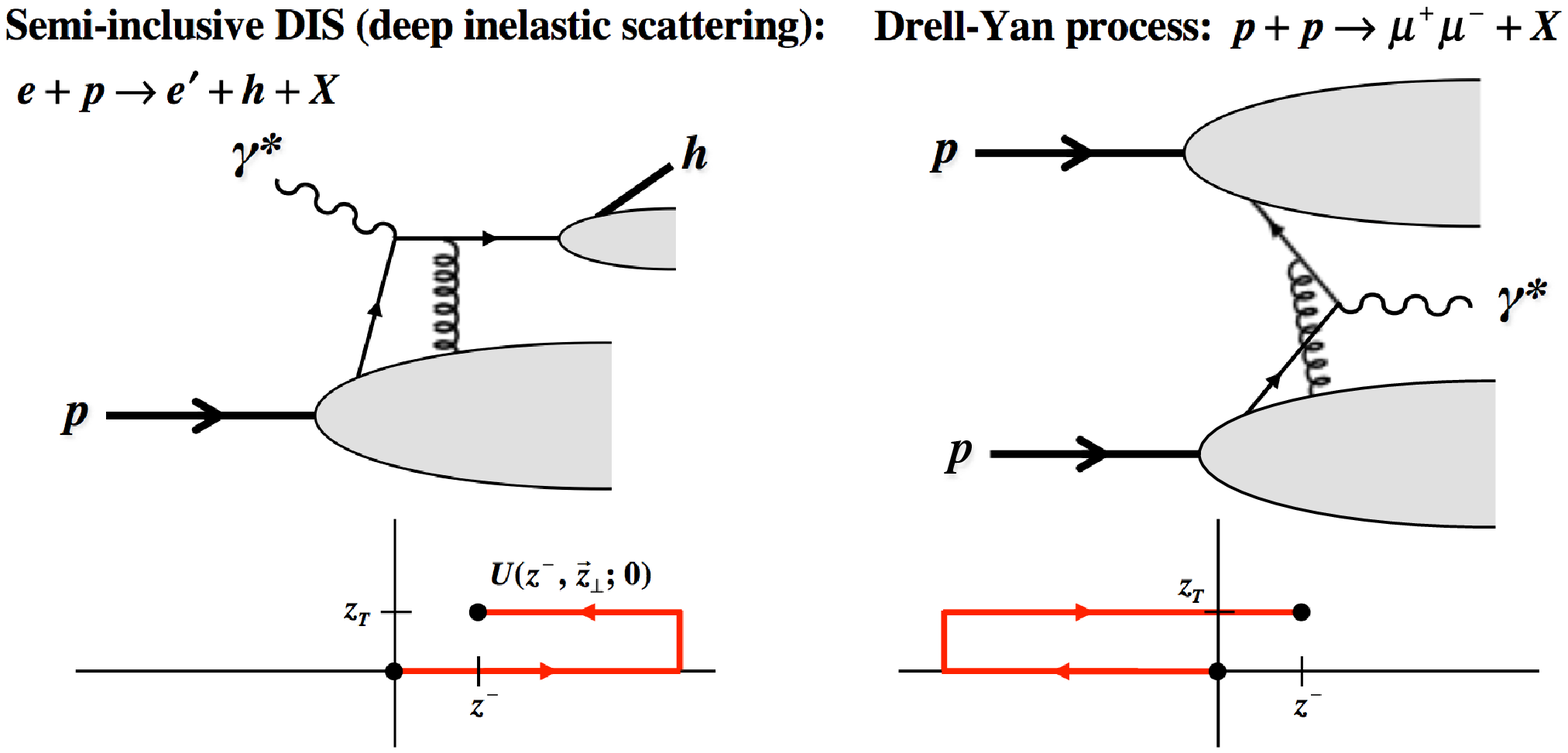}
   \end{center}
\vspace{-0.70cm}
\caption{Gauge link and color flow.}
\label{fig:c-link}
\vspace{-0.4cm}
\end{minipage} 
\hspace{-0.3cm}
\begin{minipage}[c]{0.50\textwidth}
    \vspace{-0.1cm}
   \begin{center}
     \includegraphics[width=7.5cm]{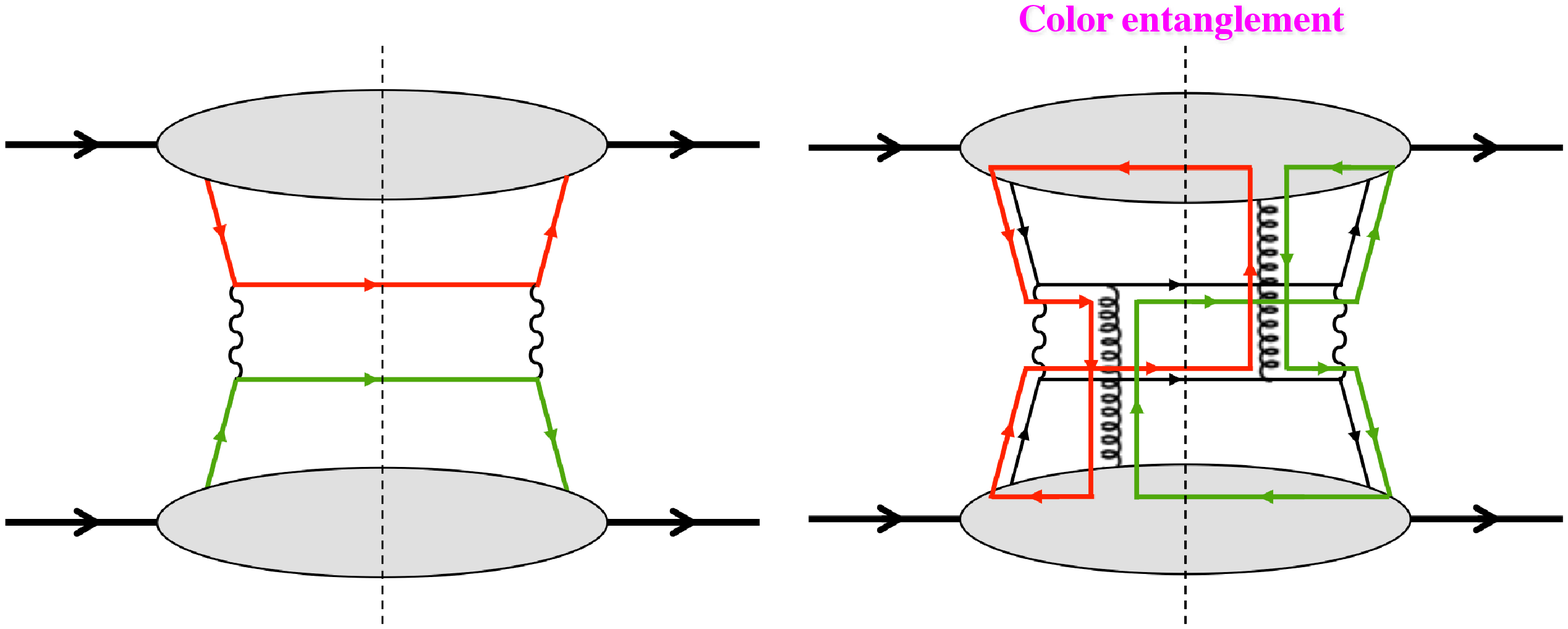}
   \end{center}
\vspace{-0.60cm}
\caption{Color entanglement.}
\label{fig:c-entanglement}
\vspace{-0.4cm}   
\end{minipage}
\end{tabular}
\vspace{0.20cm}
\end{minipage}
\end{figure}

As the TMD studies develop, color flow became important
for the first time in experiments of hadron physics. We know that the color
gauge invariance is satisfied by the intermediate gauge link
$U$ as
\begin{align}
q(x,{\rm{ }}{k_ \bot }) = 
\int {\frac{{d{z^ - }{d^2}{z_ \bot }}}{{2{{(2\pi )}^3}}}{\rm{ }}} 
{e^{ - ix{p^ + }{z^ - } + i{{\vec k}_ \bot } 
\cdot {{\vec z}_ \bot }}}
\left\langle p \right|\bar \psi ({z^ - },{\rm{ }}{\vec z_ \bot })
{\rm{ }}{\gamma ^ + }U({z^ - },{\rm{  }}{\vec z_ \bot };{\rm{ }}0)
\psi (0){\left. {\left| p \right\rangle } \right|_{{z^ + } = 0}} ,
\nonumber
\label{eqn:gauge-color}
\end{align} 
in defining the TMDs.
The color flow given by the gauge link does not play an important
role in the 1D quantities such as the longitudinal PDFs.
However, it become conspicuous in the TMDs because of additional
flow in the transverse direction as illustrated in 
Fig.\,\ref{fig:c-link}. The paths are different between
the semi-inclusive DIS and Drell-Yan, which results in
the sign difference in their TMDs.
This is the first case where the color flow appears in actual
observables, and its experimental confirmation is in progress.
It suggests an interesting future development to create a new
field with explicit color degrees of freedom.

In obtaining the PDFs, we use the experimental data which satisfy
kinematical conditions for factorization. 
For example, the cross section of the hard hadron production $p + p \to h +X$ 
in Fig.\,\ref{fig:hard-cross} is factorized into 
$\sigma = \sum_{a,b,c} f_a (x_a,Q^2) \otimes f_b (x_b,Q^2)
            \otimes \hat \sigma (a b \to c X) \otimes D_c ^ h (z,Q^2)$.
The color entanglement means that
particles interacts with each other by strong (color) interactions and
the state of each particle cannot be described independently 
from other-particle states even though the particles are separated
by a large distance. It indicates that the color entanglement should be
investigated as one of factorization-breaking effects.
Because quantum entanglement is getting popular in various fields of
science and technology, the color entanglement is an interesting
interdisciplinary topic.
With the development of TMD physics, the color entanglement became
an interesting subject although it is not easy to find such a signature.
For example, a possible factorization breaking process is shown
in Fig.\,\ref{fig:c-entanglement} for the di-jet production
$p + p \to j_1 + j_2 +X$ \cite{c-entanglement}.
The processes $p + p \to \gamma + h^\pm +X$, $\pi^0 +h^\pm +X$
were experimentally investigated to find the color entanglement;
however, it is inconclusive at this stage for such a signature.
We need much better understanding on TMD physics to push
this project in future.

\section{Summary}
\label{summary}

The unpolarized PDF studies of the nucleon have a long history of 
about a half century. Now, they became a field of precision physics 
by a wide variety of high-energy experimental measurements together 
with detailed higher-order calculations in perturbative QCD.
In addition, the field expanded to other topics, longitudinally- 
and transversely-polarized PDFs, tensor-polarized PDFs,
nuclear PDFs, fragmentation functions, and 3D structure functions
of GPDs, GDAs, and TMDs. 
Using precise PDFs, we expect that new physics beyond
the standard model will be discovered and that new quark-gluon plasma
phenomena will be found. Furthermore, we believe that the origin 
of the nucleon spin will be solved and the 3D tomography
of the nucleon will be clarified. The 3D tomography studies are 
also valuable for understanding the composition of ultra-high-energy 
cosmic rays and GZK cutoff phenomena. 
We also bear in mind that the PDF field should be developed
along with progress in other fields.
The nucleon spin and 3D tomography field is closely connected
with the gravitational physics because of the similar 
matrix elements of the angular-momentum and energy-momentum tensors.
The energy-momentum tensor part provides information on 
the gravitational form factors.
It will lead to a clarification of the origin of hadron masses 
or gravitational sources by the quark and gluon degrees of freedom.
An effect of the color gauge link appears in the TMDs, which may
be a beginning of new field on explicit color degrees of freedom.
For example, the color entanglement is an interesting phenomena
although its experimental confirmation is not easy at this stage.
With future experimental facilities such as EIC and LHeC,
we expect to have further progress of the PDF field.

\section*{Acknowledgments}

Figures 3, 7, 14 and 17 are used under 
\href{https://creativecommons.org/licenses/by/4.0/}
{the Creative Commons license} with the copyright permission of authors.
Figures 5, 6, and 11 are used with the copyright permission 
of American Physical Society and authors.
Figure 8 is used with the copyright permission of Elsevier 
with an author. Figure 15 was supplied by Tianbo Liu and Haiyan Gao.




\begin{thebibliography}{99}
\vspace{-0.20cm}
\bibitem{unpol-pdfs} R. D. Ball {\it et al.} (NNPDF Collaboration),
                       Eur. Phys. J. C {\bf 77}, 663 (2017).
\vspace{-0.20cm}
\bibitem{n3lo} For example, see 
                J. Ablinger {\it et al.}, Nucl. Phys. B {\bf 927}, 339 (2018)
                           \& {\bf 932}, 129 (2018);
                S. Moch {\it et al.}, JHEP {\bf 1710}, 041 (2017).
\vspace{-0.20cm}
\bibitem{nutev-anomaly} G. P. Zeller {\it et al.},
                   Phys. Rev. Lett. {\bf 88}, 091802 (2002);
                   S. Kumano, Phys. Rev. D {\bf 66}, 111301 (2002).
\vspace{-0.20cm}
\bibitem{lattice-PDFs} X. Ji, Phys. Rev. Lett. {\bf 110}, 262002 (2013).
    For recent progress, 
    see $e.g.$ T. Ishikawa {\it et al.}, Phys. Rev. D {\bf 96}, 094019 (2017);
    Huey-Wen Lin {\it et al.}, Prog. Part. Nucl. Phys. {\bf 100}, 107 (2018);
    Yu-Sheng Liu {\it et al.}, arXiv:1807.06566.
\vspace{-0.20cm}
\bibitem{icecube-2017} M. G. Aartsen {\it et al.} (IceCube Collaboration), 
                  Nature {\bf 551}, 596 (2017); Erratum {\bf 554}, 554 (2018).
\vspace{-0.20cm}
\bibitem{ncteq-2016} K. Kovarik {\it et al.} (nCTEQ Collaboration),
                               Phys. Rev. D {\bf 93}, 085037 (2016).
\vspace{-0.20cm}
\bibitem{j/psi-upc} V. Guzey {\it et al.}, 
                               Phys. Lett. B {\bf 726}, 290 (2013).
\vspace{-0.20cm}
\bibitem{jam-2017} J. J. Ethier, N. Sato, and W. Melnitchouk,
                          Phys. Rev. Lett. {\bf 119}, 132001 (2017).
\vspace{-0.20cm}
\bibitem{eic-2016} A. Accardi {\it et al.}, 
                          Eur. Phys. J. A {\bf 52}, 268 (2016).
\vspace{-0.20cm}
\bibitem{trans-solid-2017} Z. Ye {\it et al.} (SoLID Collaboration), 
                          Phys. Lett. B {\bf 767}, 91 (2017).
\vspace{-0.20cm}
\bibitem{jlab-b1-fermilab} Proposal to Jefferson Lab PAC-38 (PR12-11-110), 
                         J.-P. Chen {\it et al.} (2011);
      Fermilab E1039 experiment, LoI P1039 (2013);
      S. Kumano and Qin-Tao Song, Phys. Rev. D {\bf 94} (2016) 054022. 
\vspace{-0.20cm}
\bibitem{b1-conv} W. Cosyn, Yu-Bing Dong, S. Kumano, and M. Sargsian,
                          Phys. Rev. D {\bf 95}, 074036 (2017). 
\vspace{-0.20cm}
\bibitem{sk-tensor} S. Kumano, Phys. Rev. D {\bf 82}, 017501 (2010). 
\vspace{-0.20cm}
\bibitem{HKKS-2016}
    M. Hirai, H. Kawamura, S. Kumano, and K. Saito, 
    Prog. Theor. Exp. Phys. 113B04 (2016). 
\vspace{-0.20cm}
\bibitem{JAM-FFs-2016} N. Sato {\it et al.}, 
                       Phys. Rev. D {\bf 94}, 114004 (2016).
\vspace{-0.20cm}
\bibitem{unpol-tmds-2017} A. Bacchetta {\it et al.}, 
                       JHEP {\bf 06}, 081 (2017);
          I. Scimemi and A. Vladimirov,
                       Eur. Phys. J. C {\bf 78}, 89 (2018).
\vspace{-0.20cm}
\bibitem{tmd-sivers} M. Anselmino {\it et al.}, 
                          Euro. Phys. J. A {\bf 39}, 89 (2009).
\vspace{-0.20cm}
\bibitem{tmd-pretz} C. Lefty and A. Proukudin,
                          Phys. Rev. D {\bf 91}, 034010 (2015).
\vspace{-0.20cm}
\bibitem{tmd-bm} Z. Lu and I. Schmidt, Phys. Rev. D {\bf 81}, 034023 (2010);
       V. Barone, S. Melis, and A. Prokudin, 
                         Phys. Rev. D {\bf 82}, 114025 (2010).
\vspace{-0.20cm}
\bibitem{Pretzelosity-solid} H. Gao, presentation at this conference.
                    The distribution is from Ref.\,\cite{tmd-pretz}.
\vspace{-0.20cm}
\bibitem{gpds}  M. Diehl and P. Kroll,
                Euro. Phys. J. C {\bf 73}, 2397 (2013);
         D. Mueller, Few Body Syst. {\bf 55},  317 (2014); 
	     K. Kumericki, S. Liuti, and H. Moutarde,
                Eur. Phys. J. A {\bf 52}, 157 (2016);
         H. Moutarde, P. Sznajder, and J. Wagner,
                arXiv:1807.07620.
\vspace{-0.20cm}
\bibitem{j-parc-gpds} 
S. Kumano, M. Strikman, and K. Sudoh,
                          Phys. Rev. D {\bf 80}, 074003 (2009).
T. Sawada {\it et al.}, 
    Phys. Rev. D {\bf 93}, 114034 (2016) 
\vspace{-0.20cm}   
\bibitem{kk-2014} 
  H. Kawamura and S. Kumano, Phys. Rev. D {\bf 89}, 054007 (2014).
\vspace{-0.20cm}
\bibitem{kst-2018} 
  S. Kumano, Qin-Tao Song, and O. V. Teryaev,  
           Phys. Rev. D {\bf 97} (2018) 014020.
\vspace{-0.20cm}
\bibitem{counting} H. Kawamura, S. Kumano, T. Sekihara,
                       Phys. Rev. D {\bf 88}, 034010 (2013);
  Wen-Chen Chang, S. Kumano, T. Sekihara,
           Phys. Rev. D {\bf 93}, 034006 (2016).
\vspace{-0.20cm}
\bibitem{mechanical} 
V.~D.~Burkert, L.~Elouadrhiri and F.~X.~Girod,
  Nature {\bf 557} (2018) no.7705, 396; 
M.~V.~Polyakov and P.~Schweitzer, arXiv:1801.05858;   
K.~Tanaka, Phys. Rev. D {\bf 98}, 034009 (2018).
\vspace{-0.20cm}
\bibitem{pk-2010} L. Portugal and T. Kodama
                    Nucl. Phys. A {\bf 837}, 1 (2010).
\vspace{-0.20cm}
\bibitem{c-entanglement} T. C. Rogers and P. J. Mulders, 
                            Phys. Rev. D {\bf 81}, 094006 (2010);
 A. Adare $et al$. (PHENIX Collaboration), Phys. Rev. D {\bf 95}, 072002 (2017).
\vspace{-0.20cm}
\end{thebibliography}
\end{document}